\documentclass[11pt]{article}  
\usepackage{graphics}
\usepackage{vmargin} 
\setpapersize{A4}

\begin{document}

\title{Statistical Physics of the Jamming Transition: The Search for Simple Models}
\author{S. F. Edwards and D. V. Grinev\thanks{Electronic address: dg218@phy.cam.ac.uk}\\
\\Polymers and Colloids group, Cavendish laboratory,\\ University of Cambridge,\\
Madingley Road, Cambridge, CB3 0HE, UK\\}
\maketitle

\begin{center}
\Large
\textbf{Abstract}
\end{center}
\normalsize
We investigate universal features of the jamming transition in granular materials, colloids and glasses. We show that the jamming transition in these systems has common features: slowing of response to external perturbation, and the onset of structural heterogeneities.

\section{Introduction}
\label{Introduction}

Many studies in physics concern the onset of instabilities into turbulence or some form of chaotic break up, the simplest being laminar flow into turbulent flow, but the collapse of civil engineering structures is much studied. There are not many studies of the reverse phenomenon, where some chaotic  system becomes more regular, or even stops altogether.
The jamming phenomenon exists in quantum as well as classical systems, but in this article we will discuss only the latter, and indeed systems which are readily accessible to intuitive understanding.

The approach of this article will be to list the most common jamming systems, and see if there exist simple physical or mathematical systems which lead to interpretable experiments and soluble theories. In this paper, we consider mostly our own work, or studies that we know of in detail.
 The word ``jamming'' derives from systems coming to rest, and an archetypical problem is in the flow of granular or colloidal systems.
We will try to answer the following questions:
\begin{itemize}
\item What is the jammed state?
\item How does the jamming transition occur?
\item What are the common features of the jamming transition in granular
materials, colloidal suspensions and glasses?
\end{itemize}

In this paper, we confine ourselves to the discussion of the jamming transition in granular materials, colloids and glasses. We will not comment on other systems that exhibit the jamming
transition, which are many: foams, vortices in superconductors, field lines in turbulent plasma etc. All these systems throw light on one another and exhibit universal behaviour. We will demonstrate that the jamming transition in these systems can be characterized by the slowing of response to external perturbations, and the onset of structural heterogeneities.

\section{Granular Media}
\label{Stress}

\begin{figure}[h] 
\begin{center} 
\resizebox{6cm}{!}{\rotatebox{180}{\includegraphics{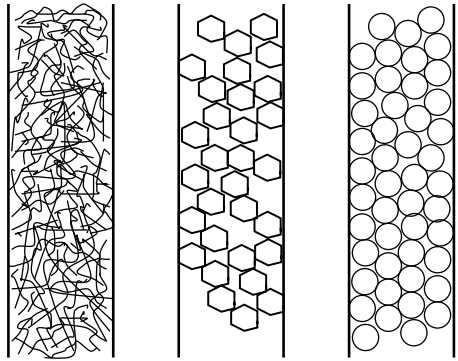}}}
\caption{Types of granular materials}
\label{column}
\end{center} 
\end{figure}

The jamming transition in dry granular media is a very common phenomenon which can be observed in everyday life. Stirring jammed sugar with a spoon can be hard but manageable. A jam within a silo can cause its failure. Nonetheless, such an ubiquity does not make the problem less difficult. In this section we will not even attempt to analyze the dynamical problem of a granular flow coming to rest, and the onset of jamming. Instead, we will try to characterize the static jammed state of a granular material. The difficulty of a theoretical analysis lies in the fact that dense granular media can expand upon shearing, a well known Reynolds dilatancy, that can be viewed as a counterpart of jamming \cite{Reynolds,Goddard}. Consider a cylindrical vertical pipe in the following cases (see Figure \ref{column}):

\begin{itemize}
\item Rough walls, filled with pieces of twisted wire. This wire will entangle and not flow at all, indeed one does not need the pipe.
\item Rough walls, with rough, approximately spherical particles. These can flow under certain conditions, but can also jam; it is a well studied problem in chemical engineering.
\item Smooth walls, smooth spheres. This is difficult, perhaps
impossible, to jam.
\end{itemize}

\begin{figure}[h] 
\begin{center} 
\resizebox{6cm}{!}{\includegraphics{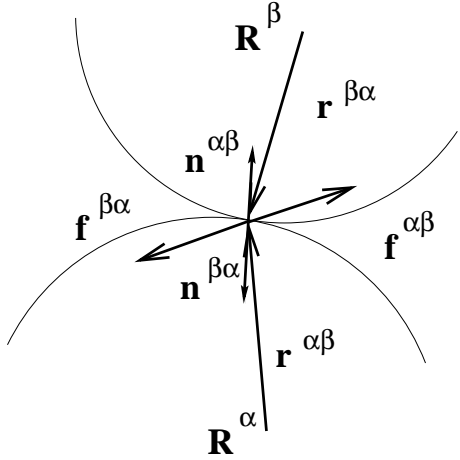}}
\caption{Detail of two grain contact}
\label{2grain}
\end{center}
\end{figure}

If one asks for the simplest granular material in the above list, it must be the third one if we model the granular material as an assembly of discrete rigid particles whose interactions with their neighbours are localized at point contacts. Therefore, the description of the network of intergranular contacts is essential for the understanding of the stress transmission and the onset of the jammed state in granular assemblies. The random geometry of a static granular packing can be visualized as a network of intergranular contacts. For any aggregate of rigid particles the transmission of stress from one point to another can occur only via the intergranular contacts. Therefore, the contact network determines the network of intergranular forces. In general, the contact network can have a coordination number varying within the system and different for every particular packing.  It follows then that the network of intergranular forces is indeterminate i.e. the number of unknown forces is larger than the number of Newton's equations of mechanical equilibrium. Thus, in order for the network of intergranular forces to be perfectly defined, the number of equations must equal the number of unknowns. This can be achieved by choosing the contact network with a certain fixed coordination number. In this case the system of equations for intergranular forces has a unique solution and the complete system of equations for the stress tensor can be derived. This is the simplest model of a granular material. If this cannot be solved, one will be left with empiricism. This is from the point of view of physics; it is not a good philosophy for engineers. The geometric specification of our system is as follows: we will need $z$ contact point vectors  $\vec{{\cal R}}^{\,\alpha\beta}$, centroid of a grain $\alpha$, $\vec{R}^{\,\alpha}$, $\vec{r}^{\,\alpha\beta}$ the vector from $\vec{R}^{\,\alpha}$ to $\vec{{\cal R}}^{\,\alpha\beta}$, and the distance between grains $\alpha$ and $\beta$, $\vec{R}^{\,\alpha\beta}$ (see Figure \ref{2grain}).

 Grain $\alpha$ exerts a force on grain $\beta$ at a point $ \vec{{\cal R}}^{\,\alpha\beta}\,=\,\vec{R}^{\,\alpha}+\vec{r}^{\,\alpha\beta}$. The contact is a point in a plane whose normal is $\vec{n}^{\,\alpha\beta}$. The vector $\vec{R}^{\,\alpha}$ is defined by:

\begin{equation}
\vec{R}^{\,\alpha}=\frac{\sum_{\,\beta} \,\vec{{\cal R}}^{\,\alpha\beta}}{z}\, ,
\label{conpoint}
\end{equation}
so that $\vec{R}^{\,\alpha}$ is the centroid of contacts, and hence the relation: $\sum_{\beta} \vec{r}^{\,\alpha\beta}\,=\,0$. We note that $z$ is the number of contacts per grain, and $\sum_{\,\beta}$ means summation over the nearest neighbours. We define the distance between grains $\alpha$ and $\beta$
\begin{equation}
\vec{R}^{\,\alpha\beta}=\vec{r}^{\,\beta\alpha}- \vec{r}^{\,\alpha\beta}\, ,
\label{centroid}
\end{equation}
 Hence $\vec{R}^{\,\alpha}$, $\vec{r}^{\,\alpha\beta}$ and $\vec{n}^{\,\alpha\beta}$ are geometrical properties of the aggregate under consideration and the other shape specifications do not enter the equations. In a static array, Newton's equations of intergranular force and torque balance are satisfied. Balance of force around the grain requires

\begin{equation}
\sum_{\beta}f^{\,\alpha\beta}_{i}=g^{\alpha}_{i}\, ,
\label{newt2}
\end{equation}

\begin{equation}
f^{\,\alpha\beta}_{i}+f^{\,\beta\alpha}_{i}=0\, ,
\label{newt3}
\end{equation}
where ${\vec g}^{\,\alpha}$ is the external force acting on grain $\alpha$.

The equation of torque balance is

\begin{equation}
\sum_{\beta}\epsilon_{ikl}\,f^{\,\alpha\beta}_{k}\,r^{\,\alpha\beta}_{l}=C_{i}^{\,\alpha}\, .
\label{couple}
\end{equation}
Friction is assumed to be infinite \cite{Tabor}. It can be verified that, for the intergranular forces in the static array to be determined by these equations,  the coordination number $z=3$ in 2-D and $z=4$ in 3-D is required. The microscopic version of stress analysis is to determine all of the intergranular forces, given the applied force and torque loadings on each grain, and geometric specification of a granular array. The number of unknowns per grain is $zd/2$. The required force and torque equations give $d + \frac{d(d-1)}{2}$ constraints. The system of equations for the intergranular forces is complete when the coordination number is  $z_m \,=\,d+1$. In addition, the configuration of contacts and normals is only acceptable if all the forces are compressive. If tensile forces are allowed, then we would be studying a sandstone (an interesting problem, but not a subject for a study of jamming). Many investigators do not believe that $z$ will be this small, and would invoke the kind of arguments used in bridgework, where the same problem arises when extra spares lead to too few equations for solution. However the simplest assumption is to assume that if the geometry gives more than $\frac{z_m}{2}$ contacts, some will contain no force. At all events we can restrict ourselves to systems where there really are $\frac{z_m}{2}$ contacts per grain The main question is: can one observe the jammed state in this simple model  where the static stress state can be determined?
 We define the tensorial force moment:

\begin{equation}
S^{\,\alpha}_{ij}=\sum_{\beta}f^{\,\alpha\beta}_{i}\,r^{\,\alpha\beta}_{j}\, ,
\label{microstress}
\end{equation}
which is the microscopic analogue of the stress tensor. With $C_{i}^{\,\alpha}=0$, $S^{\,\alpha}_{ij}$ will be symmetric. To obtain the macroscopic stress tensor from the tensorial force moment, we coarse-grain, i.e. average it over an ensemble of configurations:

\begin{equation}
\sigma_{ij}({\vec r})\,=\,\langle \,\sum_{\alpha=1}^{N}\,S_{ij}^{\,\alpha}\, \delta({\vec r}-{\vec R^{\,\alpha}})\,\rangle\, .
\label{macrostress}
\end{equation}
The number of equations required equals the number of independent  components of a symmetric stress tensor $\sigma_{ij}\,=\,\sigma_{ji}$, and is $\frac{d\,(d\,+\,1)}{2}$. At the same time, the number of equations available is $d$. These are vector equations of the stress equilibrium  $\frac{\partial \sigma_{ij}}{\partial x_{j}}\,=\,g_{i}$ which have their origin in Newton's second law. Therefore we have to find  $\frac{d(d-1)}{2}$ equations, which possess the information from Newton's third law, to complete and solve the system of equations which governs the transmission of stress in a granular array.

Given the set of equations $(\ref{newt2}-\ref{couple})$ we can write the probability functional for the integranular force  $f_{ij}^{\,\alpha\beta}$ as

\begin{eqnarray}
 P \left \{ f_{i}^{\,\alpha\beta} \right \} &=&{\cal N}
\delta\big(\sum_{\beta}f^{\,\alpha\beta}_{i}-g^{\alpha}_{i}\big) \nonumber \\
& &\mbox{}\times
\delta\big(\sum_{\beta}\epsilon_{ikl}\,f^{\,\alpha\beta}_{k}\,r^{\,\alpha\beta}_{l}\big) \nonumber \\
& &\mbox{}\times
\delta\big(f^{\,\alpha\beta}_{i}+f^{\,\beta\alpha}_{i}\big)\,,
\label{forcepdf}
\end{eqnarray}
where the normalization, ${\cal N}$, which is a function of a configuration, is

\begin{eqnarray}
{\cal N}^{-1}&=&\int\prod_{\alpha,\beta}\, P \left \{f_{i}^{\,\alpha\beta} \right\}\,{\cal D}f^{\,\alpha\beta}\, .
\label{norm}
\end{eqnarray}

The probability of finding the tensorial force moment $S_{ij}^{\,\alpha}$ on grain $\alpha$ is

\begin{eqnarray}
P \left \{ S_{ij}^{\,\alpha} \right \} &=& \int\prod_{\alpha,\beta}
\delta\Big (S^{\,\alpha}_{ij}-\sum_{\beta}f^{\,\alpha\beta}_{i}r^{\,\alpha\beta}_{j}\Big ) \,P \left \{ f_{i}^{\,\alpha\beta} \right \} \,{\cal D}f^{\,\alpha\beta}\, ,
\label{stresspdf}
\end{eqnarray}
where $\int\,{\cal D}f^{\,\alpha\beta}$ implies integration over all functions $f^{\,\alpha\beta}$, since all the constraints on $f^{\,\alpha\beta}$ have been experienced. We assume that the $z=d+1$ condition means that the integral exists.

It has been shown \cite{Grinev1,Grinev2,Ball} that the fundamental equations of stress equilibrium take the form

\begin{equation}
\nabla_{j}\,\sigma_{ij}\,+\,\nabla_{j}\nabla_{k}\nabla_{m}\,K_{ijkl}\sigma_{lm}+...\,=\,g_{i}
\label{stressforceeq}
\end{equation}

\begin{equation}
 P_{ijk}\sigma_{jk} + \nabla_{j}T_{ijkl}\sigma_{kl} + \nabla_{j}\nabla_{l}U_{ijkl}\sigma_{km} + ... =0\, .
\label{stressgeom}
\end{equation}
However, there are difficulties with the averaging procedure \cite{Grinev1} which is highly non-trivial.
The simplest mean-field approximation gives the equation $\sigma_{11}\,=\,\sigma_{22}$ for the case of an isotropic and homogeneous disordered array \cite{Grinev1}. Though this equation gives the diagonal stress tensor $\sigma_{ij}\,=\,p\,\delta_{ij}$, it is not rotationally invariant. The only linear, algebraic and rotationally invariant equation is $\mbox{Tr}\,\sigma_{ij}\,=\,0$. However, this equation cannot be accepted, for a stable granular aggregate does not support tensile stresses. We believe that, at least in the simplest case, the fundamental equation for the microscopic stress tensor should be linear and algebraic (because of the linearity of Newtons' second law for intergranular forces). In this paper we offer an alternative way which is considered below.

\begin{figure}[h] 
\begin{center} 
\resizebox{3cm}{!}{\includegraphics{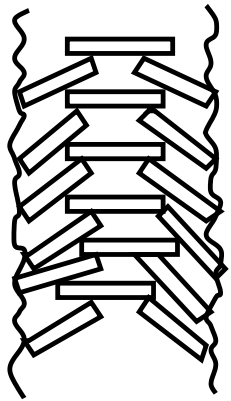}}
\caption{Jammed pipe}
\label{jpipe}
\end{center}
\end{figure}

The leading terms of the system of equations (\ref{stressforceeq},\ref{stressgeom}) arise from the system of discrete linear equations for $S_{ij}^{\,\alpha}$ \cite{Grinev1,Grinev2}

\begin{equation}
\sum_{\beta}\,S_{ij}^{\,\alpha}M_{jk}^{\,\alpha}R^{\,\alpha\beta}_{k}\,-\,S_{ij}^{\,\beta}M_{jk}^{\,\beta}R_{k}^{\,\beta\alpha}\,=\,g_{i}^{\,\alpha}
\label{sfesimpl}
\end{equation}

\begin{equation}
S^{\,\alpha}_{11}\,-\,S^{\,\alpha}_{22}\,=\,2S^{\,\alpha}_{12}\,\mbox{tan}\,\theta^{\,\alpha}
\label{sgesimpl}
\end{equation}
and if $\mbox{tan}\,\theta^{\,\alpha}$ has an average value $\mbox{tan}\,\phi$

\begin{equation}
\sigma_{11}\,-\,\sigma_{22}\,=\,2\sigma_{12}\,\mbox{tan}\,\phi
\end{equation}
which is known as the Fixed Principal Axes equation \cite{Bouchaud,Wittmer1,Wittmer2,Claudin}, and has been used with notable effect to solve the problem of the stress distribution in sandpiles. For a homogeneous and isotropic system, the averaging process gives the stress tensor $\sigma_{ij}\,=\,p\,\delta_{ij}$ which is simply hydrostatic pressure, as is to be expected. Rotating $S_{ij}^{\alpha}$ by some arbitrary angle $\theta$ one can easily that (\ref{sgesimpl}) constraints the off-diagonal components of $S^{\alpha}_{ij}$ to be zero. The system of equations (\ref{sfesimpl},\ref{sgesimpl}) is solved by Fourier transformation and the macroscopic stress tensor is obtained by averaging over the angle $\theta$

\begin{equation}
i\sigma_{11}({\vec k})\,=\,\langle\,S_{11}({\vec k})\,\rangle_{\theta}\,=\,\frac{g_{1}(k_{1}^{3}\,+\,3k_{2}^{2}k_{1})\,+\,g_{2}(k_{2}^{3}\,-\,k_{1}^{2}k_{2})}{|{\vec k}|^{4}}
\end{equation}

\begin{equation}
i\sigma_{22}({\vec k})\,=\,\langle\,S_{22}({\vec k})\,\rangle_{\theta}\,=\,\frac{g_{2}(k_{2}^{3}\,+\,3k_{1}^{2}k_{2})\,+\,g_{1}(k_{1}k_{2}^{2}\,-\,k_{1}^{3})}{|{\vec k}|^{4}}
\end{equation}

\begin{equation}
i\sigma_{12}({\vec k})\,=\,\langle\,S_{12}({\vec k})\,\rangle_{\theta}\,=\,\frac{(g_{1}k_{2}-g_{2}k_{1})(k_{2}^{2}\,-\,k_{1}^{2})}{|{\vec k}|^{4}}
\label{avsol3}
\end{equation}
where $|{\vec k}|^{2}\,=\,k_{1}^{2}\,+\,k_{2}^{2}$ and $\sigma_{ij}({\vec r})\,=\,\int\,\sigma_{ij}({\vec k})\,e^{i{\vec k}{\vec r}}\,\mbox{d}^{3}{\vec k}$. 
By doing the inverse Fourier transformation one can see that the macroscopic stress tensor is diagonal.
There must also be constraints on the permitted configurations (due to the absence of tensile forces) which are not so easily expressed, for they affect each grain in the form

\begin{equation}
S^{\,\alpha}_{ik}\,M^{\,\alpha}_{kl}\,R^{\,\alpha\beta}_{l}\,n^{\,\alpha\beta}_{i}\,>\,0
\label{microfric}
\end{equation}
which has not yet been put into continuum equations other than $\mbox{Det}\,\sigma\,>\,0$ and $\mbox{Tr}\,\sigma\,>\,0$. However, this condition must be crucial, for without it jamming becomes
a very ubiquitous phenomenon provided the density of grains is such that they are all in contact.
This example shows the utility of simple models. Apply this type of grain in a vertical pipe (the grain is a thick tile, see Figure \ref{jpipe}). 

This system is clearly jammed, but the problem is how it can get into this configuration. Another limiting case is to think of a sphere with many spines pointing in the radial direction. These spines mean that spheres can only approach or retreat along the direction joining their centres, and the only other motion permitted to a group of two or more is that of rigid rotation of the group. As soon as a line (in 2-D) or shell (3-D) of these objects occurs, they jam; so they always jam. Again, although this is a possible material, most materials fall into the classes above. So there are trivial jamming problems, but it is proving difficult to produce an effective analytic theory for intermediate materials. It is natural then to use computer simulations, and there is a notable literature in existence \cite{Goddard}. However, rather than comment on this literature
for granular materials we move to the related problem of colloidal flow which offers a natural route from grains to glasses.

\section{Colloidal Suspensions}
\label{Colloids}

The simplest model of a colloid  which exhibits the jamming transition is the sheared suspension of hard monodisperse spheres, interacting hydrodynamically through a Newtonian solvent of viscosity $\eta_{0}$. Such a system at equilibrium is characterized by its volume fraction $\phi_{v}$. The behaviour of this system at equilibrium is well known\cite{Russel}. With increasing $\phi_{v}$, the system crystallizes, with phase
 coexistence occurring between $\phi_{v}\,\approx\,0.50$ and $\phi_{v}\,\approx\,0.55$. If simple shear is applied, we need one other parameter, which is the Peclet number
\begin{equation}
\mbox{Pe}\,=\,\frac{6\,\pi\,\dot{\gamma}\,\eta_{0}\,d^{3}}{k_{B}\,T}
\label{peclet}
\end{equation}
where $d$ is the particle diameter, $\dot{\gamma}$ the shear rate and $T$ the temperature. The Peclet number gives the ratio of shear forces to Brownian forces. Pairwise hydrodynamic interactions between the neighbouring particles can be divided into squeeze terms along the line of centres and terms arising from relative shear motion. To leading order the squeeze hydrodynamic force on a particle is given by the well-known Reynolds lubrication formula

\begin{equation}
f_{i}\,=\,-\,\sum_{j}\,\frac{3\,\pi\,\eta_{0}}{8\,h_{ij}}\,\Big\{(v_{i}\,-\,v_{j})\cdot n_{ij}\Big\} \,n_{ij}\,+\,O\Big(\ln\frac{2}{h_{ij}}\Big)
\end{equation}
where the sum is over nearest-neighbour particles $j$, $h_{ij}$ is the gap between the neoghbour surfaces with the unit of distance the particle diameter, $n_{ij}$ is the unit vector along the line of centres $i$ to $j$ and $v_{i}$, $v_{j}$ are the particle centre velocities.

In the absence of other interactions, Brownian forces are left to control the approach of particles. If conservative (steric or charge) forces are present, these control the gaps between particles, and dominate over the Brownian forces at high shear rates. When one studies the shear flow of such a system, it exhibits thickening effects: a rise of viscosity with increasing shear rate (for a review, see \cite{Barnes}). At volume fractions approaching random close
packing, $\phi_{RCP}\,\approx\,0.64$, discontinuous thickening with a large jump in viscosity occurs at a critical $\mbox{Pe}$. However, at a lower  $\phi_{v}$ and lower $\mbox{Pe}$ a more continuous rise can be observed (see the Figure \ref{jamming} below). Hence, the jamming transition in this simple system can be either continuous or discontinuous, which depends sensitively on the volume fraction. 

\begin{figure}[h] 
\begin{center}
\resizebox{5cm}{!}{\rotatebox{-90}{\includegraphics{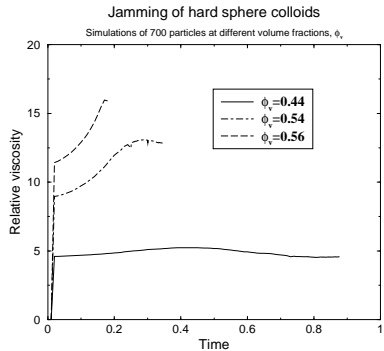}}}
\caption{Relative viscosity of hard sphere colloids versus time, for various volume fractions}
\label{jamming}
\end{center}
\end{figure}

\begin{figure}[h]
\begin{center}
\resizebox{5cm}{!}{\includegraphics{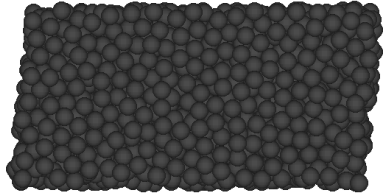}}
\caption{Snapshot from a 3-D simulation of 2000 hard spheres at $\phi_{v}\,=\,0.56$. All particles are shown.}
\label{whole}
\end{center}
\end{figure}

\begin{figure}[h]
\begin{center}
\resizebox{5cm}{!}{\includegraphics{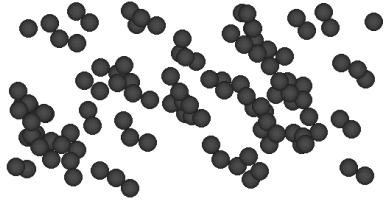}}
\caption{Snapshot from a 3-D simulation of 2000 hard spheres at $\phi_{v}\,=\,0.56$. Only particles who have separations $h\,<\,10^{-4}\,d$ are shown. The flow direction is left to right.}
\label{cluster}
\end{center}
\end{figure}

This happens in various experimental systems but in particular, in those whose particles, by polymer coating or surface charges, do not flocculate (if they want to stick together then jamming is not an obvious phenomenon). Colloids with repulsive interactions exhibit shear thinning i.e. the viscosity decreases as the shear rate increases. The presence of aggregating forces can greatly alter the shear thinning. We will discuss the regime of shear thickening (which we call later the jamming transition) in the simplest model, gradually incorporating various interactions. The physical picture can be obtained by combining theory and computer simulations\cite{Melrose0,Melrose,Melrose1,Farr,Farr1,Farr2,Catherall}. The simulation technique for particles under quasi-static motion determined by a balance of conservative and dissipative forces has been proposed in \cite{Melrose0}. The motion of $N$ colloidal particles, immersed within a hydrodynamic medium, is governed by an equation of quasi-static force balance

\begin{equation}
-{\mathcal R}({\vec X})\,\cdot \,{\vec V}\,+\,{\vec F}_{C}({\vec X})\,+\,{\vec F}_{B}(t)\,=\,{\vec{0}}
 \label{qsfbe}
\end{equation}
where ${\vec X}$ represents the $6N$ particle position coordinates and orientations, ${\mathcal R}$ is a $6N\,\times\,6N$ resistance matrix and ${\vec V}\,=\,d{\vec X}/dt$ is the particle velocity. The terms $F_{C}$ and $F_{B}$ represent conservative and Brownian forces. The effects of inertia are ignored i.e. the Reynolds number is small

\begin{equation}
\mbox{Re}\,=\,\frac{\rho_{s}\,\dot{\gamma}\,d^{2}}{\eta_{0}}\,\ll\,1
\label{reynolds}
\end{equation}
Bearing in mind that for a typical colloid $\frac{\mbox{Pe}}{\mbox{Re}}\,\sim\,10^{7}$, it follows that it is possible to achieve $\mbox{Re}\,\ll\,1$ and $\mbox{Pe}\,\gg\,1$ simultaneously.  Equation (\ref{qsfbe}) can be solved on a computer\cite{Melrose0,Melrose}. Although simulation shows exactly what is happening in the sense one can see every sphere flowing its path (see Figures \ref{whole},\ref{cluster}), it is still not agreed as to whether the phenomenon of thickening is an order-disorder transition\cite{Hoffman1,Hoffman2} or whether it is due to the development of clusters of particles along the compression axis\cite{Brady,Butera}. 

A microscopic kinetic theory for the origin of the increased bulk viscosity at high $\phi_{v}$ and $\mbox{Pe}\,=\,\infty$ has been recently proposed \cite{Farr,Farr2}. It attributes the viscous enhancement to the presence of hydrodynamic clustering, i.e. incompressible groups of particles which lie near the compression axis. The rigidity of clusters is provided by divergent lubrication drag coefficients. The theory predicts a critical $\phi_{v}$ above which jamming occurs. This model gives a flow-jam phase transition at any strain and may be of more general applicability (e.g. colloids with conservative and Brownian forces). Assuming the cluster length to be additive on collision, one can obtain the standard Smoluchowski aggregation equation

\begin{equation}
\frac{\mbox{d}\,X_{k}}{\mbox{d}\,t}\,=\,\frac{1}{2}\,\sum_{i,j=1}^{\infty}\,\Big[K_{ij}\,X_{i}\,X_{j}\,-\,X_{i+j}\Big]\,\Big(\delta_{i+j,k}\,-\,\delta_{i,k}\,-\,\delta_{j,k}\Big)
\label{smoleq}
\end{equation}
This equation relies upon a mean field approximation, i.e. it is assumed that each cluster is embedded in an average flow, composed of the other clusters, "spectator" particles and solvent. Eqn (\ref{smoleq}) governs the evolution of the concentrations $X_{k}$ of clusters of size $k$ monomers as a function of time, given that clusters can collide, aggregate and break up at rates dependent on their sizes and specified by the aggregation kernel $K_{ij}$. The mathematical procedure of solving Equation (\ref{smoleq}) has been reported in \cite{Farr,Farr2}. We discuss the generic structure of
the theory. The rheology of this system results from a competition between a binary aggregation process and single cluster breakup. The paradigm of hydrodynamic clustering modelled in terms of aggregation-breakup laws provides the qualitative physical picture of the jamming transition. At low $\phi_{v}$ breakup dominates. The system achieves a steady state with a population of clusters whose average size increases and a viscosity which increases as the volume fraction rises. In the vicinity of some volume fraction $\phi_{c}$, average cluster size diverges, and so does the viscosity

\begin{equation}
\eta\,\propto\,\eta_{0}\,\Big(1\,-\,\frac{\phi_{v}}{\phi_{c}}\Big)^{-2}
\label{visco2}
\end{equation}
Above $\phi_{c}$ the flow is transient and dominated by aggregation. The average cluster size diverges after a certain strain, which falls as $\phi_{v}$ is further increased. The system undergoes the jamming transition and can never reach a steady state. This model predicts $\phi_{c}\,\approx\,0.49$, although experiments suggest a divergence of bulk viscosity at a higher volume fraction (which is still less than $\phi_{RCP}$). This difference is due to the presence of conservative interactions in the real hard core colloid.
 In conclusion, we briefly discuss the jamming behaviour of long flexible chain polymers. A good way to understand this is to plot viscosity against the molecular weight of the polymer in the molten state. The molecular weight is equivalent to the length of the chain. The short chains slide past each other in many ways, giving a linear relation between length and shear viscosity, but an entanglement crisis occurs at a critical molecular weight $M_{c}$, when the polymer can only wriggle in Brownian motion up and down and ``out'' of a tube formed by its neighbours in reptation. Whether there is sharp transition is not established in a difficult slow experiment, but all agree that the dependence of the viscosity jumps to $M^{3.4}$, a colossal change which at
first sight is just like a jamming, but very slowly the melt flows flows due to reptation. Polymers very easily form glasses and it only takes a fall in temperature to make the melt solidify, although the specific heat shows that although reptation (a slow translatory motion) has ceased, there is plenty of other movement taking place, but movement which averages over time to zero. A further fall in temperature destroys this motion, and one enters a state like a conventional glass.

\section{Glasses}
\label{Glasses}

Studies of glasses seem to fall into two camps. Those taking continuum field representation of the material and solving under statistical thermodynamics conditions, principally mode-coupling methods, and alternatively the translation, into comparatively simple equations and simple physical models. The former has given intuitive thinkers a hard time because it is not obvious in those cases, such as the polymeric glasses discussed below, where one should have a picture of what goes, what indeed is going on. The modelling approach is weak; for example, in an assembly of packed spheres where an enormous number of motions is possible, but strong in polymer glasses where the motion is obvious and the jamming of motion is the cessation of the centre of mass diffusion. The mode-coupling theory\cite{Gotze} of the glass transition for simple liquids has an ergodicity to non-ergodicity transition. It starts with an equation of motion for the density correlation function, which is an integro-differential Langevin equation, with a non-linear memory kernel. This non-linear memory term governs the transition to the non-ergodic state, and can be characterized by measurements of the dynamic structure factor. We refer the reader to the literature \cite{Gotze} for details of this approach. The glass transition is not limited to special types of materials. Every class of material can be transformed in an amorphous solid if the experimental parameters are adjusted to the dynamics of the system. Consider therefore, two extreme examples. Consider the system consisting of spherical molecules such as rare gases. The hopping time of the spheres is very short, and the dynamics extremely fast. Nevertheless, it has been shown that such fluids undergo a glass transition from the super cooled melt, if one cools the system with a quenching rate of $q\,\propto\,10^{12}\, \mbox{Ks}^{-1}$, when the molecules jam effectively to rest. On the other hand, consider polystyrene which consists of very large molecules. The dynamics of such a system is much more complicated in comparison to spheres because there are many degrees of freedom. It is most significant that the centre of mass diffusion of a single molecule is small.  We want to stress characteristics of the glass transition in general.
 The most significant points we want to discuss are:

\begin{itemize}
\item The divergence of the transport or inverse transport properties, such as viscosity, inverse diffusion coefficient, and relaxation times.
\item The extreme broad relaxation phenomena of the stress, modulus etc.
\item The quantitative definition of the term cooperativity.
\item Influence of external parameters on $T_{g}$.
\end{itemize}
It has been recognised that the relaxation time follows the Vogel -- Fulcher Law (VF):
\begin{equation}
\tau\,\propto\,e^{\frac{A}{T-T_{0}}} \label{VF}
\end{equation} 

\begin{figure}[h]
\begin{center} 
\resizebox{6cm}{!}{\includegraphics{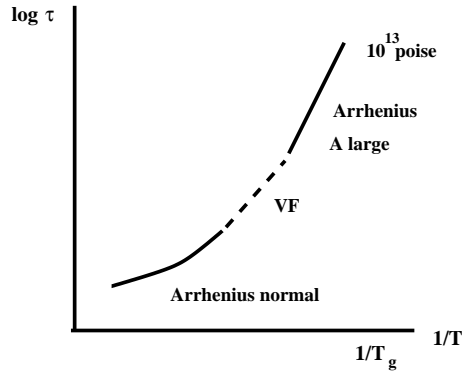}}
\caption{Vogel-Fulcher Law}
\label{VFfig}
\end{center}
\end{figure}

\begin{figure}[h]
\begin{center} 
\resizebox{6cm}{!}{\includegraphics{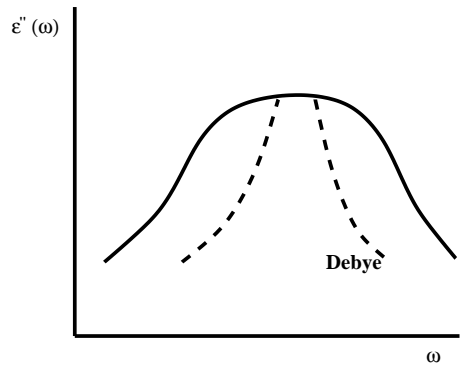}}
\caption{Kohlrausch-Williams-Watts Law}
\label{KWWfig}
\end{center}
\end{figure}

This law has many names. In polymer physics it is often called the Williams -- Landel -- Ferry (WLF) law \cite{Ferry}. The law is not valid over the whole temperature range (see Figure \ref{VFfig}). The divergence given by (\ref{VF}) comes at $T_{0}$, where $T_{0}$ is a temperature below the freezing temperature $T_{g}$, and the empirical rule is given by:

`\begin{equation}
T_{0}\,\simeq\,T_{g}\,-\,(20\,\sim\,30)^{0}. \label{er}
\end{equation}
The physical meaning of $T_{g}$ is still unclear, and we want to try to clarify this point. The VF law is much stronger than the critical slowing down in phase transition phenomena where, by scaling arguments, the relaxation time is given by

\begin{equation}
\tau\,\sim\,\mid T\,-\,T_{0}\mid^{-\nu z} \label{rt}
\end{equation}
where $\nu$ is the correlation length exponent i.e. $\xi\,\sim\,(T-T_{0})^{-\nu}$ and $z$ the dynamical exponent

\begin{equation}
\tau\,\sim\,\xi^{z}\, . \label{dexp}
\end{equation}
There have been attempts to fit data for freezing transitions (glass transition, spin glass transition) by Eqn.(\ref{rt}), and it turned out that $\nu z$ is very high $\nu z \,\sim\, 10,...,20$ which seems very unnatural. This again indicates a physical significance to the VF law. Another peculiar point lies in the relaxation properties. An empirical law  was found long ago by Kohlrausch and in the early seventies recovered by  Williams and Watts in their studies of the broadening of relaxation processes \cite{Williams}. For example, measurement of the dielectric constant shows a much larger half width in the imaginary part compared to the Debye process. Empirically this is described by the Kohlrausch -- Williams -- Watts (KWW) law (see Fig \ref{KWWfig})

\begin{equation}
\phi(t)\,\sim\,e^{-(\frac{t}{\tau})^{\,\beta}} \label{KWW}
\end{equation}
where $\phi(t)$ can be any quantity which relaxes, i.e.

\begin{equation}
\phi(t)\,=\,\frac{\epsilon (t)\,-\,\epsilon (\infty)}{\epsilon(0)\,-\,\epsilon (\infty)} \label{dcr}
\end{equation}
However, there has been no convincing explanation for a unique value for $\beta$. We doubt that there is more to the physics of (\ref{KWW}) than a wide relaxation spectrum due to different physical processes, and a more relevant question is how both the KWW and VF laws link together in general, and what is the relationship to cooperativity. It is believed that the glass transition exhibits a large amount of cooperative motion as the system is close to $T_{g}$. This might be indicated as well by the VF law, which is the crossover to the Arrhenius behaviour right at $T_{g}$ but with an extremely high activation energy (see Figure \ref{VFfig}). This activation energy is so high that it could hardly be attributed to only one molecule, and the phenomenological interpretation is that there are cooperative regions of some linear size diverging at some temperature

\begin{equation}
\xi\,\sim\,(T\,-\,T_{g})^{power}\, . \label{cdiv}
\end{equation}
Another quite general question is the state of time-temperature superposition principle. This says that if one measures a physical quantity, $D(t)$, at some temperature $T$, and if the measurement is repeated at some temperature $T_{1}$, the quantity $D(t)$ can be resolved by a ``shift factor'' $a_{T}$, i.e. $D(T,t)$ is not a function of two variables but only of a combination of both:

\begin{equation}
D(T,t)\,=\,D \Big(\frac{t}{a_{T}} \Big) \label{Dfun}
\end{equation}
where $a_{T}$ is often given by

\begin{equation}
a_{T}\,=\,\frac{C_{1}\,+\,T_{g}}{T\,-\,T_{g}\,+\,C_{2}}
\label{afun}
\end{equation}
near $T_{g}$, and

\begin{equation}
a_{T}\,=\,\frac{\delta E}{T}\,-\,\frac{\delta E}{T^{*}}
\label{afun2}
\end{equation}
far from $T_{g}$. Hence (\ref{afun}) is of the Volger-Fulcher form.

\begin{figure}[h]
\begin{center} 
\resizebox{6cm}{!}{\includegraphics{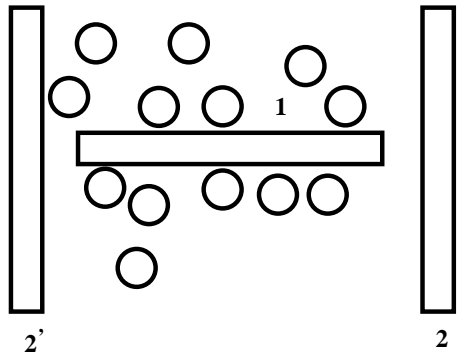}}
\caption{``Jammed'' rod}
\label{rodgt}
\end{center}
\end{figure}

Polymeric glasses offer two challenges:

\begin{itemize}
\item There is a clear intuitive picture of what is happening. The ``tube'' closes at its ends, or contracts at ``entanglement'' points.
\item The molecular weight offers, as with viscosity, a new degree of freedom, and hence new laws emerge. Any theory of glass must encompass VF, KWW, and the experiments we now describe.
\end{itemize}
For polymers, the glass transition temperature depends on the molecular weight, and an empirical rule is given by Flory and Fox:

\begin{equation}
T_{g}(L)\,=\,T_{g}(\infty)\,-\,\frac{\mbox{constant}}{L}\, ,
\label{emrule}
\end{equation}
where $L$ is the length of the molecules i.e. the molecular weight. The agreement of (\ref{emrule}) with experiments is not particularly good, but it gives an estimate of $T_{g}(L)$. Hence (\ref{emrule}) tells us that there is not a very significant dependence on the molecular weight, unless $L$ is small. This point should be investigated using the knowledge of polymer dynamics in melts which has recently emerged. Another empirical law we want to discuss is the mixing rule in plastification. Mixing two glass forming polymers together, the new $T_{g}$ is often given by

\begin{equation}
\frac{1}{T_{g}(\mbox{mix})}\,=\,\frac{\phi_{1}}{T_{g_{1}}}\,+\,\frac{\phi_{2}}{T_{g_{2}}}
\label{emrulemix}
\end{equation}
to zeroth order. The $\phi_{i}$ is the volume fraction of the $i$'th polymer. 
We have discussed so far the most important experimental results. Clearly there is a need for a solvable model in the framework of which the VF and KWW laws can be derived.  
Any fundamental theory of glass transition should relate the mobility of a molecule within the cage to mobilities of molecules forming the cage. These mobilities of surrounding molecules are coupled with those of their neighbours {\em etc.}, and therefore in general there is no small parameter that can justify the decoupling approximation. 
The small molecule model of a glass transition involves cages, but polymers are simpler because they have a tube, and a tube of such a model can be modelled as a straight tube, indeed restrictions along the tube can be included, but are not here for simplicity. The reason why we take rods was the advantage of simple geometry, no internal degrees of freedom, and very slow dynamics, so that we have no problems with high quenching rates, so we do not have to worry about thermodynamics. It is obvious that the small width to the length ratio and the sufficiently large number of neighbouring rods justify the decoupling procedure. If the solution of the rods is dense, the concentration $c\,\leq\,d^{2}\,L$ ($d$ is the diameter of the rod, $L$ its length and $c$ the concentration) and severe constraints are acting on the rods. For example, they cannot move rotationally, and they can only make progress along their length. Such a solution has been called entangled. Suppose now we have such a solution of highly entangled rods. A rod can slide between the entangling rods until it meets rods which block it (see Figure \ref{rodgt}).

The motion of a rod will then be like a particle diffusing along a line but meeting gates which open and close randomly through thermal fluctuations. If no barriers we present the probability $P(x,t)$ of finding the test rod at $x$ (which is the coordinate of the rod down the tube) and time $t$ satisfies the simple diffusion equation

\begin{equation}
\Big(\frac{\partial}{\partial t}\,-\,D_{0}\,\frac{\partial^{2}}{\partial x^{2}}\Big)\,P(x,t)\,=\,0
\end{equation}

\begin{figure}[h]
\begin{center} 
\resizebox{6cm}{!}{\includegraphics{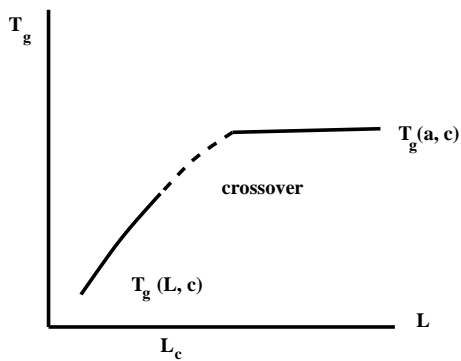}}
\caption{$T_{g}$ as a function of $L$}
\label{GTL}
\end{center}
\end{figure}

This has the solution

\begin{equation}
P(x,t)\,=\,\int_{-\infty}^{\infty}\,G_{0}(x,x';t,t')\,P_{0}(x',t')\,\mbox{d}x'
\end{equation}
where $P_{0}(x',t')$ is the initial probability function and $G_{0}(x,x';t,t')$ is the standard Green function of the diffusion equation.
Suppose now , a reflecting barrier is placed at position $R$ at time $t_{R}$ and removed at some time $t_{Q}$. Then using the method of images to find the Green function it is straightforward to calculate $P(x,t)$ \cite{Evans}. The same method can be applied when there are many barriers appearing and disappearing along the path of the rod.   
After neglecting correlations between barriers one can see that the rod (1) can only diffuse if

\begin{equation}
D\,=\,D_{0}(1\,-\,\alpha) \label{dif}
\end{equation}
where $\alpha\,\sim\,\epsilon\,(c\,D\,L^{2})^{\frac{3}{2}}$, $c\,D\,L^{2}$ being the Onsager number. Eqn.(\ref{dif}) gives $D\,=\,0$ if $\alpha\,=\,1$ i.e. $\epsilon\,c\,D\,L^{2}\,=\,1$. This is the jammed state. The result can be modified to include cooperativity. The complete solution has been obtained in \cite{Evans,Vilgis} by mapping the problem onto the self-avoiding walk problem. The VF law appears when summed over $n$ (where $n$ is the number of rods that loops are made out of)

\begin{equation}
D\,\sim\,D_{0}\,exp\,\Big(-\frac{\alpha_{1}^{2}}{1\,-\,\alpha_{2}}\Big)
\label{VF2}
\end{equation}
where the parameter $\alpha_{1}$ contains generalised constants and a minimum loop size, and $\alpha_{2}\,=\,\alpha\,-\,\alpha_{1}$. It can be shown that the number of rods moving cooperatively in the loops are given by

\begin{equation}
\bar{n}\,\sim\,(1\,-\,\alpha_{2})^{-2} \label{nof}
\end{equation}
and the size of the loop is therefore

\begin{equation}
\xi^{2}\,\sim\,\bar{n}\,L^{2}\,\sim\,\frac{L^{2}}{(1\,-\,\alpha_{2})^{2}}
\label{sol}
\end{equation}
Thus $\xi$ diverges if $\alpha_{2}\,\rightarrow\,1$, as the phenomenological interpretation requires.
So far we have used $\alpha$ to denote the expression $cdL^{2}$, a combination of constants well known in liquid crystal theory since the work of Onsager. However, its role in our theory is much more general; for example, $d$ can be temperature dependent through the fact that Van der Waals forces appear in the form $e^{-\frac{E}{nT}}$, whereas hardcore forces do not contain $T$. Thus, at the level of a model one can regard $\alpha\,=\,1$ at $T\,=\,T_{g}$. Until one studies a detailed physical case, there is no purpose in false verisimilitude. Thus the above model now says  that the VF law is a direct consequence of cooperativity. Turning now to the relaxation behaviour, it can be shown that for the rod model, the stress relaxation follows, to first order, the law

\begin{equation}
\sigma(t)\,\sim\,e^{-(\frac{t}{\tau})^{\,\frac{1}{2}}}
\label{srel}
\end{equation}
Cooperativity as indicated by (\ref{VF2}) does not change (\ref{srel}) drastically, giving rise to logarithms

\begin{equation}
\sigma(t)\,\sim\,e^{-(\frac{t}{\tau})^{\,\frac{1}{2}}}\,\Big(1\,+\,\mbox{const.}\,\mbox{log}\,\frac{t}{\tau}\Big)
\label{srel2}
\end{equation}
This leads us to the conclusion that cooperativity is responsible for the VF behaviour, but not for the relaxation phenomena. Further study of this model allows a derivation of (\ref{emrule},\ref{emrulemix}). A final interesting feature of the tube model is that the tube itself must be a random walk of step length $a$ (in the rheology literature this is called the primitive path) and the ``rod'' of the preceding discussion is of length $a$ for a very long polymer, and $L$ for a long, but not very long molecule. The freezing of large scale motion is the freezing of motion on the scale of the primitive path step length, and the long term reptative motion of the whole chain is not a vital constitutive of the glass temperature. Hence $T_{g}$ is a function of $a$ and the density of the material, and also additional parameters (e.g. chain stiffness), i.e. $T_{g}\,=\,T_{g}\Big(a(c,L), ...\Big)$. But $a$ is a function of the density itself, so that the effect of a diluent acts on both $a$ and $c$, and we expect a different concentration dependence of $T_{g}$ above $M_{c}$. $T_{g}$ dependence on $L$ is roughly given by Figure \ref{GTL}.

This simple model has been modified for arrays of rigid rods with fixed centres of rotation \cite{Obukhov}.
The rods attached to the sites of the cubic lattice can rotate freely but cannot cross each other. It is important to note that the glass transition in this system is decoupled from the structural transition (nematic ordering) and the only parameter is the ratio of the length of the rods to the distance between the centers of rotation. The Monte-Carlo study in 2-D shows that with increasing the parameter of the model a sharp crossover to infinite relaxation times can be observed. In 3-D the simulation gives a real transition to a completely frozen state at some critical length $L_{c}$. 

\subsection{An Important Analogy}

\begin{figure}[h]
\begin{center} 
\resizebox{6cm}{!}{\rotatebox{-90}{\includegraphics{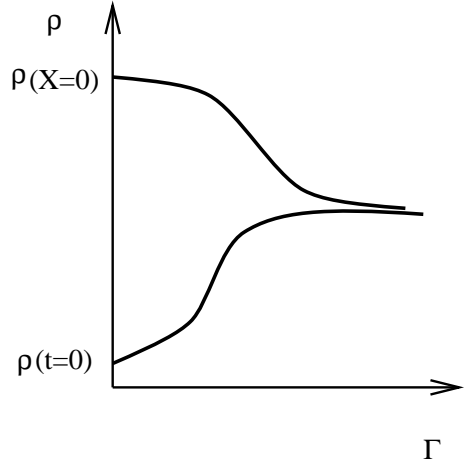}}}
\caption{Dependence of the steady-state packing density on the tapping history (Nowak et al.). Experimental values of density packing fraction are in the following correspondence with  model parameters: $\rho(X=0)=\frac{1}{v_{0}}\approx 0.64, \rho(t=0)=\rho_{0}=\frac{1}{v_{1}}\approx 0.58$ and $\rho(X=\infty)=\frac{2}{(v_{0}+v_{1})}\approx 0.62$. The vibration intensity is parametrized by $\Gamma\,=\,\frac{a}{g}$}
\label{data}
\end{center}
\end{figure}

Recent crucial experiments on granular materials \cite{Nowak1,Nowak2} show that external vibrations lead to a slow approach of the packing density to a final steady-state value. Depending on the initial conditions and the magnitude of the vibration acceleration, the system can either reversibly move between steady-state densities, or can become irreversibly trapped into metastable states; that is, the rate of compaction and the final density depend sensitively on the history of vibration intensities that the system experiences (see Figure \ref{data}).

The function which has been found to fit the ensemble averaged density $\rho(t)$ better than other functional forms, is \cite{Nowak2}:

\begin{equation}
\rho(t)=\rho_{f}\,-\,\frac{\Delta\,\rho_{f}}{1\,+\,B\,\log(1+\frac{t}{\tau})}
\label{loglaw}
\end{equation}
where the parameters $\rho_{f}$, $\Delta\,\rho_{f}$ and $\tau$ depend only on $\Gamma$.
 This is an analogue of a glass forming material where the lower curve corresponds to the situation where the quenching speed inflicted on the glass is faster than the speed at which the glass relaxes back to equilibrium. The special feature of the granular material is that it has no power of its own to proceed to the equilibrium state, so that every aspect can be studied without having to worry about the fact that the true glass is always seeking equilibrium. It is worth considering a simple model for the density response to external vibrations \cite{Grinev3}. If we assume that all configurations of a given volume are equally probable, it is possible to develop the formalism \cite{Edwards3,Mehta2,Edwards4,Edwards5,Edwards6,Higgins} analogous to conventional  statistical mechanics. We introduce the  volume function $W$ (the analogue of a Hamiltonian) which depends on the coordinates of the grains, and their orientations. Averaging over all possible configurations of the grains in real space gives us a configurational statistical ensemble, which describes the random packing of grains.  An analog of the microcanonical probability distribution is:

\begin{equation}
 P=e ^{-\frac{S}{\lambda}}\delta(V-W) \, .
\label{microcan}
\end{equation}
We can define the analogue of temperature as:

\begin{equation}
 X=\frac{\partial V}{\partial S}.
\label{compac}
\end{equation}
 This fundamental parameter is called compactivity \cite{Edwards3}.  It characterizes the packing of a granular material, and may be interpreted  as being characteristic of the number of ways it is  possible to arrange the grains within the system into a volume $\Delta V$, such that the  disorder is $\Delta S$. Consequently, the two limits of $X$ are $0$ and $\infty$,  corresponding to the most and least compact stable arrangements.  This is clearly a valid parameter for sufficiently dense powders, since one can in principle calculate the configurational entropy of an arrangement of grains, and therefore derive the compactivity from the basic definition. We will use the canonical probability distribution

\begin{equation}
 P=e ^{\frac{Y-W}{\lambda X}},
\label{gibbsdis}
\end{equation}
where $\lambda$ is a constant which gives the entropy the dimension of volume.
We call $Y$ the effective volume; it is the analogue of the free energy:

\begin{equation}
 e ^{-\frac{Y}{\lambda X}}=\int \,e ^{-\frac{W(\mu)}{\lambda X}}\,\mbox{d (all)}, \qquad
\label{effvol}
\end{equation}

\begin{equation}
V=Y-X \frac{\partial Y}{\partial X}. \label{vol}
\end{equation}
Examples of volume functions for particular systems can be found elsewhere\cite{Edwards4,Edwards5,Edwards6,Higgins}. We consider the rigid grains powder dominated by friction deposited in a container which will be shaken or tapped (in order to consider the simplest case, we ignore  other possible interactions, e.g. cohesion, and do not distinguish between the grain-grain interactions in the bulk and those on the boundaries). We assume that most of the particles in the bulk do not acquire any  non ephemeral kinetic energy, i.e. the change of a certain configuration occurs due to continuous and cooperative rearrangement of a free volume between the neighbouring grains. The simplest  volume function is

\begin{equation}
 W=v_{0}+(v_{1}-v_{0})(\mu_{1}^2+\mu_{2}^2)
\label{vfmu}
\end{equation}
where two degrees of freedom  $\mu_{1}$ and $\mu_{2}$ define the ``interaction'' of a grain with its nearest neighbours. If we assume that all grains in the bulk experience the external vibration as a random force, with zero correlation time, then the process of compaction can be seen as the Ornstein-Uhlenbeck process for  the degrees of freedom $\mu_{i},\,i=1,2$ \cite{Risken}. Therefore we write the Langevin equation:

\begin{equation}
  \frac{d \mu_{i}}{dt} + \frac{1}{\nu}\frac{\partial W}{\partial \mu_{i}}\,=\,\sqrt{D}\,f_{i}(t)
\label{langeq}
\end{equation}
where $\langle f_{i}(t)f_{j}(t')\rangle=2\delta_{ij}\delta(t-t')$, and $\nu$ characterizes the frictional resistance imposed on the grain by its nearest neighbours. The term $f_{i}(t)$ on the RHS of (\ref{langeq}) represents the random force generated by a tap. The derivation of this gives the analogue of the Einstein relation that $\nu=(\lambda X)/D$. If we identify $f$ with the amplitude of the force $a$ used in the tapping, the natural way to make this dimensionless is to write the ``diffusion'' coefficient  as :

\begin{equation}
D=\Big(\frac{a}{g}\Big)^{2}\,\frac{\nu\omega^{2}}{v}\, ,
\label{difcoeff}
\end{equation}
that is we have a simplest guess for a fluctuation-dissipation relation:

\begin{equation}
\lambda X=\Big(\frac{a}{g}\Big)^{2}\,\frac{\nu^{2}\omega^{2}}{v}
\label{fdt}
\end{equation}
where $v$ is the volume of a grain, $\omega$ the frequency of tapping, and $g$ the gravitational acceleration. The standard treatment of the Langevin equation (\ref{langeq}) is to use it to derive the Fokker-Planck equation:

\begin{equation}
 \frac{\partial P}{\partial t} =\Big( D_{ij}\frac{\partial^{2}}{\partial \mu_{i}\partial \mu_{j}}+\gamma_{ij}\frac{\partial}{\partial\mu_{i}}\mu_{j} \Big)P=0 \label{fpeq}
\end{equation}
where $D_{ij}=D\delta_{ij}$ and $\gamma_{ij}=\gamma\delta_{ij}$. This equation can be solved explicitly \cite{Grinev3,Risken}. We can calculate the volume of the system as a function of time and compactivity, $V(X,t)$. Though this is a simple model, it is too crude to give a quantitative agreement with experimental data; however, it gives a clear physical picture of what is happening. It is possible to imagine an initial state where all the grains are improbably placed, i.e. where each grain has its maximum volume $v_{1}$. So if one could put together a powder where the grains were placed in a high volume configuration, it will just sit there until shaken; when shaken, it will find its way to the distribution (\ref{gibbsdis}). It is possible to identify physical states of the powder  with characteristic values of volume in our model. The value $V=Nv_{1}$ corresponds to the ``deposited'' powder, i.e. the powder is put into the most unstable condition possible, but friction holds it.   When $V=Nv_{0}$ the powder is shaken into closest packing. The intermediate value of $V=(v_{0}+v_{1})/2$ corresponds to the minimum density of the reversible curve. Thus we can offer an interpretation of three values of density presented in the  experimental data \cite{Nowak1}.  Our theory gives three points $\rho(X=0), \rho(X=\infty)$ and $\rho(t=0)$ which are in the ratio: $v_{0}^{-1},\,\frac{2}{(v_{0}+v_{1})},\,v_{1}^{-1}$ and these are in reasonable agreement with experimental data: $\rho(X=0)=\frac{1}{v_{0}}\approx 0.64,\,\rho_{0}=\frac{1}{v_{1}}\approx 0.58$ and$\rho(X=\infty)=\frac{2}{(v_{0}+v_{1})}\approx 0.62$. Another important issue is the validity of the  compactivity concept for a ``fluffy'', but still mechanically stable, granular array, e.g. for those composed of spheres with $\rho \le 0.58$. In our theory, $\rho(X=\infty)$ corresponds to the beginning of the reversible branch (see Figure \ref{data}). We foresee that granular materials will throw much light onto  glassy behaviour in the future, for questions like the pressure fluctuations in the jammed material are accessible in granular materials, and not entirely in glasses.

\section{Discussion}

In Section \ref{Stress} we have given the analysis of stress for the case of granular arrays with the fixed values of the coordination number. However, a realistic  granular packing can have a fluctuating coordination number which need not necessarily be 3 in 2-D or 4 in 3-D. How one can extend the formalism of Section \ref{Stress} to make it capable of dealing with arbitrary packings? The simplest (however sensible) idea is to assume that in 3-D the major force is transmitted only through {\em four} contacts (we call them active contacts). The rest of the contacts transmit only an infinitesimal stress and can be christened as passive. A considerable experimental evidence for this conjecture does exist. Photoelastic visualization experiments \cite{Jaeger,Liu} show that stresses in static granular media concentrate along certain well-defined paths. A disproportionally large amount of force is transmitted via these stress paths within the material. Computer simulations \cite{Thornton2,Thornton3,Radjai1,Radjai2} also confirm the existence of well-defined stress-bearing paths. At the macroscopic scale, the most obvious characteristic of a granular packing is its density. As it has been shown in Section \ref{Glasses} it is natural to introduce the volume function $W$ and compactivity $X\,=\,\frac{\partial V}{\partial S}$. The statistical ensemble now includes now the set of topological configurations with different microscopic force patterns. Therefore, in order to have a complete set of physical variables, one should combine the volume probability density functional (\ref{microcan}) with the stress probability density functional (\ref{stresspdf}).
The joint probability distribution functional can be written in the form

\begin{equation}
 P \left \{ V \mid S_{ij}^{\,\alpha} \right \} \,=\,e ^{-\frac{S}{\lambda}}\,\delta(V-W)\,\prod_{\alpha,\beta}
\delta\Big (S^{\,\alpha}_{ij}-\sum_{\beta}f^{\,\alpha\beta}_{i}r^{\,\alpha\beta}_{j}\Big )\,\Theta(\vec{f}^{\,\alpha\beta} \,\cdot\, \vec{n}^{\,\alpha\beta})\, P \left \{ \vec{f}^{\,\alpha\beta} \right \} .
\label{microcangen}
\end{equation}
We speculate that this mathematical object is necessary for the analysis of the stress distribution in granular aggregates with an arbitrary coordination number. However, this is an extremely difficult problem which involves a mathematical description of force chains i.e. the mesoscopic clusters of grains carrying disproportionally large stresses and surrounded by the sea of spectator particles. An explicit theoretical analysis of this problem  will be a subject of future research.

The aim of this paper has been to show that simple models can capture the basic physics of the jamming transition and provide insight into universal features of this phenomenon. We have christened the "jamming transition" as diverse physical phenomena which take place in granular materials, colloids and glasses. However, we argue that the existence of similar features like the slowing of response and the appearance of heterogeneous correlated structures gives hope that there are universal physical laws that govern jamming in various systems. Due to the extreme complexity of this problem, the simplest models should be considered and solved at first. This would help to capture basic mechanisms of these phenomena, and establish a theoretical framework which will incorporate higher complexities and details, and become a predictive tool.

\section{Acknowledgements}

We acknowledge financial support from Leverhulme Foundation (S. F. E.) and Shell (Amsterdam) (D. V. G.). The authors have benefited from many conversations with Professors Robin Ball, Mike Cates, Joe Goddard, Sidney Nagel and Tom Witten. We are most grateful to Dr John Melrose and Alan Catherall for providing the diagrams and many illuminating discussions on the jamming transition in colloids. We thank the Institute for Theoretical Physics (University of California in Santa-Barbara) for hosting the ``Jamming and Rheology'' programme, and for warm hospitality.

\end{document}